\documentclass[superscriptaddress,showpacs,preprintnumbers,preprint,nofootinbib]{revtex4-1}

\usepackage{amsmath,amssymb,graphicx}

\begin{document}

\title{Spontaneous leptonic CP violation and nonzero $\theta_{13}$}
\date{\today}

\author{G.~C.~Branco}
\email{gbranco@ist.utl.pt}
\affiliation{Departamento de F\'{\i}sica and Centro de F\'{\i}sica Te\'{o}rica de Part\'{\i}culas, Instituto Superior T\'{e}cnico, Universidade T\'ecnica de Lisboa, Av. Rovisco Pais, 1049-001 Lisboa, Portugal}

\author{R. Gonz\'{a}lez Felipe}
\email{ricardo.felipe@ist.utl.pt}
\affiliation{Departamento de F\'{\i}sica and Centro de F\'{\i}sica Te\'{o}rica de Part\'{\i}culas, Instituto Superior T\'{e}cnico, Universidade T\'ecnica de Lisboa, Av. Rovisco Pais, 1049-001 Lisboa, Portugal}
\affiliation{Instituto Superior de Engenharia de Lisboa, Rua Conselheiro Em\'{\i}dio Navarro, 1959-007 Lisboa, Portugal}

\author{F.~R.~Joaquim}
\email{filipe.joaquim@ist.utl.pt}
\affiliation{Departamento de F\'{\i}sica and Centro de F\'{\i}sica Te\'{o}rica de Part\'{\i}culas, Instituto Superior T\'{e}cnico, Universidade T\'ecnica de Lisboa, Av. Rovisco Pais, 1049-001 Lisboa, Portugal}

\author{H.~Ser\^{o}dio}
\email{hserodio@cftp.ist.utl.pt}
\affiliation{Departamento de F\'{\i}sica and Centro de F\'{\i}sica Te\'{o}rica de Part\'{\i}culas, Instituto Superior T\'{e}cnico, Universidade T\'ecnica de Lisboa, Av. Rovisco Pais, 1049-001 Lisboa, Portugal}

\begin{abstract}
We consider a simple extension of the Standard Model by adding two Higgs triplets and a complex scalar singlet to its particle content. In this framework, the CP symmetry is spontaneously broken at high energies by the complex vacuum expectation value of the scalar singlet. Such a breaking leads to leptonic CP violation at low energies. The model also exhibits an $A_4 \times Z_4$ flavour symmetry which, after being spontaneously broken at a high-energy scale, yields a tribimaximal pattern in the lepton sector. We consider small perturbations around the tribimaximal vacuum alignment condition in order to generate nonzero values of $\theta_{13}$, as required by the latest neutrino oscillation data. It is shown that the value of $\theta_{13}$ recently measured by the Daya Bay Reactor Neutrino Experiment can be accommodated in our framework together with large Dirac-type CP violation. We also address the viability of leptogenesis in our model through the out-of-equilibrium decays of the Higgs triplets. In particular, the CP asymmetries in the triplet decays into two leptons are computed and it is shown that the effective leptogenesis and low-energy CP-violating phases are directly linked.
\end{abstract}

\maketitle

\section{Introduction}

The solid evidence for neutrino oscillations and, consequently, for nonzero neutrino masses and mixing, has established leptogenesis~\cite{Fukugita:1986hr,Davidson:2008bu} as one of the most appealing mechanisms to explain the matter-antimatter asymmetry observed today in our universe. The most attractive feature of this mechanism relies on the fact that the interactions relevant for leptogenesis can simultaneously be responsible for the non-vanishing and smallness of neutrinos masses, once the well-known seesaw mechanism is invoked. Among the canonical seesaw realizations, the so-called triplet (or type II) seesaw~\cite{typeIIseesaw} is probably the most economical in the sense that it counts with a single source of flavour structure, namely, the symmetric complex Yukawa coupling matrix $\mathbf{Y}^\Delta$ that couples the $SU(2)_L$ scalar triplet $\Delta$ to leptons. Furthermore, in its minimal realization, with only one scalar triplet, the flavour pattern of $\mathbf{Y}^\Delta$ uniquely determines the flavour structure of the low-energy effective neutrino mass matrix $\mathbf{m}_\nu$. In this regard, it is worth recalling that the current solar and atmospheric neutrino oscillation data are consistent with the so-called tri-bimaximal leptonic mixing pattern~\cite{Harrison:2002er}, which from the theoretical point of view seems to call for a discrete family symmetry. Along this line many models have been proposed, with a vast majority relying on the $A_4$ symmetry (for recent reviews on flavour symmetries see e.g. Refs.~\cite{Altarelli:2010gt,Ishimori:2010au}).

Unfortunately, leptogenesis cannot be successfully implemented in the minimal triplet seesaw scenario. The CP asymmetry induced by the triplet decays is generated beyond the one-loop level and, consequently, is highly suppressed. Thus, new sources for neutrino masses are necessary to induce the required lepton asymmetry via the out-of-equilibrium triplet decays~\cite{Ma:1998dx,Hambye:2000ui,Hambye:2003ka,Hambye:2003rt,DAmbrosio:2004fz}. In the latter case, estimates of the thermal leptogenesis efficiency~\cite{Hambye:2000ui,Hambye:2003ka} as well as a more precise calculation of it by solving the full set of Boltzmann equations~\cite{Hambye:2005tk} indicate that successful leptogenesis typically requires the lightest triplet mass to be large enough, $M_\Delta \gtrsim 10^{9}$~GeV, if no extra sources of CP violation are present and lepton flavour effects are not taken into account.

A crucial ingredient of any dynamical mechanism which aims at explaining the baryon asymmetry is the violation of the CP symmetry. For scalar triplet leptogenesis, the complex Yukawa couplings of the Higgs triplets to leptons, as well as their complex couplings to the standard Higgs doublet, usually provide the necessary source of (explicit) CP violation. Alternatively, the required amount of CP violation can be generated if CP is spontaneously broken by the complex vacuum expectation value (VEV) of a scalar field. Besides being more attractive from a theoretical viewpoint, the latter framework is also more economical since both the CP violation necessary to generate the baryon asymmetry and leptonic CP violation~\cite{Branco:2011zb} potentially observable at low energies come from a common source, namely, from the phase of the scalar field responsible for the spontaneous CP breaking at a high energy scale~\cite{Branco:2003rt}.

The purpose of this work is to present a simple scenario where the above three aspects (leptogenesis, leptonic mixing and spontaneous CP violation) are related. To this aim, we shall add to the Standard Model (SM) a minimal particle content: two Higgs triplets $\Delta_a\, (a=1,2)$ with unit hypercharge and a complex scalar singlet $S$ with zero hypercharge. In our framework, neutrinos acquire masses via the well-known type II seesaw mechanism, implemented by the two scalar triplets, and leptogenesis becomes viable due to the out-of-equilibrium decays of the latter in the early universe. Furthermore, we shall assume that the CP symmetry is spontaneously broken at high energies by the complex VEV of the singlet $S$, leading to the CP violation required for leptogenesis as well as to low-energy leptonic CP violation. By imposing an $A_4 \times Z_4$ flavour symmetry which is spontaneously broken at a high scale with a specific vacuum configuration, the tribimaximal (TBM) lepton mixing is obtained. Nonzero values of $\theta_{13}$, required by the neutrino oscillation data from T2K~\cite{Abe:2011sj} and MINOS~\cite{Adamson:2011qu}, arise once we consider small perturbations around the vacuum alignment configuration which leads to exact TBM mixing. We show that in our framework the value of $\theta_{13}$ measured by the Daya Bay Reactor Neutrino Experiment~\cite{DayaBay} can be obtained, leading to large Dirac-type CP violation. Finally, we discuss how leptogenesis is realized in order to show that sizable CP asymmetries can be obtained. Moreover, the effective leptogenesis phase can be directly linked to the low-energy leptonic CP-violating phases.

\section{The model: particle content and symmetries}

Let us consider the SM extended with two Higgs triplets $\Delta_a\, (a=1,2)$ of unit hypercharge and a complex scalar singlet $S$ with zero hypercharge. In the $SU(2)$ representation:
\begin{align}
\Delta_a=\begin{pmatrix}
         \Delta^0_a & -\Delta^+_a/\sqrt{2} \\
         -\Delta^+_a/\sqrt{2} & \Delta^{++}_a \\
       \end{pmatrix}.
\end{align}

We impose CP invariance at the Lagrangian level and introduce a $Z_4$ symmetry under which the scalar and lepton fields transform as follows:
\begin{align} \label{Z4sym}
S \rightarrow -S, \quad \phi \rightarrow i \phi, \quad \Delta_1 \rightarrow \Delta_1, \quad \Delta_2 \rightarrow -\Delta_2, \quad L_j \rightarrow i L_j, \quad e_{Rj} \rightarrow -i e_{Rj},
\end{align}
where $\phi = (\phi^+\, \phi_0)^T$ is the standard Higgs doublet of $SU(2)_L$; $L_j$ and $e_{Rj}$ $(j=1,2,3)$ are the SM lepton doublets and right-handed singlets, respectively. The remaining fields transform trivially under the $Z_4$ symmetry. As it turns out, the above symmetry assignment yields a quite simple and predictive scenario.

The most general scalar potential invariant under the above symmetries can be written as
\begin{align}\label{pot}
    V^{CP\times Z_4}=V_S+V_\phi+V_\Delta+V_{S\phi}+V_{S\Delta}+V_{\phi \Delta}+V_{S\phi \Delta},
\end{align}
where
\begin{align}\label{potterms}
\begin{split}
V_S&=\mu_S^2\,(S^2+S^{\ast 2}) + m_S^2\,S^\ast S+\lambda_S^\prime\,(S^4+S^{\ast 4})+\lambda_S^{\prime\prime} S^\ast S (S^{2}+ S^{\ast 2})+\lambda_S\left(S^\ast S\right)^2,\\
V_\phi&=m_\phi^2\,\phi^\dagger \phi +\lambda_\phi\,(\phi^\dagger \phi)^2,\\
V_\Delta&=\displaystyle\sum\nolimits_{a} M^2_a\,\text{Tr}(\Delta_a^\dagger \Delta_a)+ \displaystyle\sum\nolimits_{a,b} \left[(\lambda_{\Delta})_{ab} \text{Tr}(\Delta_a^\dagger \Delta_a)\,\text{Tr}(\Delta_b^\dagger \Delta_b)+(\lambda^\prime_{\Delta})_{ab} \text{Tr}(\Delta_a^\dagger \Delta_a \Delta_b^\dagger \Delta_b)\right],\\
V_{S\phi}&= \eta_S\,(S^\ast S) (\phi^\dagger \phi)+ \eta_S^\prime\left(S^2+S^{\ast 2}\right) (\phi^\dagger \phi),\\
V_{S\Delta}&=\displaystyle\sum\nolimits_{a}\text{Tr}\left(\Delta_a^\dagger \Delta_a\right)\left[\eta_a\,(S^2+\,S^{\ast 2})+\xi_a\,S^\ast S\right]\,,\\
V_{\phi \Delta}&= \displaystyle\sum\nolimits_{a}\left[\xi^\prime_{a}\,(\phi^\dagger\phi)\text{Tr}\left(\Delta_a^\dagger \Delta_a\right)+ \xi^{\prime\prime}_{a}\,(\phi^\dagger\Delta_a^\dagger \Delta_a \phi)\right] + (\mu_2 M_2 \tilde{\phi}^T\Delta_2\tilde{\phi} + \text{H.c.}) ,\\
V_{S\phi \Delta}&=\tilde{\phi}^T\Delta_1\tilde{\phi}\left(\lambda_{1}\,S +\lambda^\prime_{1}\,S^\ast\right)+\text{H.c.},
\end{split}
\end{align}
and $\tilde{\phi} = i \sigma_2 \phi^\ast$. Since CP invariance has been imposed at the Lagrangian level, all the parameters are assumed to be real. Notice that the choice of the $Z_4$ symmetry given in Eq.~(\ref{Z4sym}) forbids the term $\mu_1 M_1 \tilde{\phi}^T \Delta_1 \tilde{\phi}$ which is crucial for the mechanism of leptogenesis to be viable in the present model. Yet, once the singlet $S$ acquires a complex VEV, this term is generated from the scalar potential contribution $V_{S\phi \Delta}$ in Eq.~(\ref{potterms}).

In order to generate a realistic lepton mixing pattern we shall also impose an $A_4$ discrete symmetry at high energies (see discussion in Section~\ref{sawneut}). We recall that, in a particular basis, the Clebsch-Gordan decompositions of the $A_4$ group are $\mathbf{1^\prime} \otimes \mathbf{1^{\prime \prime}}=\mathbf{1}$
and $\mathbf{3}\otimes \mathbf{3}=\mathbf{1}\oplus\mathbf{1^\prime}\oplus\mathbf{1^{\prime\prime}}\oplus \mathbf{3_s}\oplus \mathbf{3_a}$. Moreover, the Clebsch-Gordan coefficients for the product of two triplet fields with components $a_{1,2,3}$ and $b_{1,2,3}$ are given by
\begin{align}
\begin{split}
&(a \otimes b)_\mathbf{1}=a_1 b_1 + a_2 b_3 + a_3 b_2,\\
&(a \otimes b)_\mathbf{1^\prime}=a_3 b_3 + a_1 b_2 + a_2 b_1,\\
&(a \otimes b)_\mathbf{1^{\prime\prime}}=a_2 b_2 + a_1 b_3 + a_3 b_1,\\
&(a \otimes b)_\mathbf{3_s}=\frac{1}{3}(2a_1 b_1 - a_2 b_3 - a_3 b_2,2a_3 b_3 - a_1 b_2 - a_2 b_1,2a_2 b_2 - a_1 b_3 - a_3 b_1),\\
&(a \otimes b)_\mathbf{3_a}=\frac{1}{2}(a_2 b_3 - a_3 b_2,a_1 b_2 - a_2 b_1,a_1 b_3 - a_3 b_1),
\end{split}
\end{align}
while the symmetric product of three triplets reads
\begin{align}
\begin{split}
((a \otimes b)_{\mathbf{3_s}} \otimes c)_\mathbf{1}=&\frac{1}{3}\left(2a_1 b_1 c_1 + 2a_2 b_2 c_2 + 2a_3 b_3 c_3 - a_1 b_2 c_3 - a_1 b_3 c_2\right.\\
&\left.-a_3 b_1 c_2-a_3 b_2 c_1-a_2 b_1 c_3-a_2 b_3 c_1\right).
\end{split}
\end{align}

The spontaneous breaking of the $A_4$ symmetry is then guaranteed by adding to the theory two extra heavy scalar fields, $\Phi$ and $\Psi$, with a suitable VEV alignment. The complete symmetry assignments of the fields under $A_4 \times Z_4$ and $SU(2)_L \times U(1)_Y$ are given in Table~\ref{reps}.

\begin{center}
\begin{table}[t]
\caption{\label{reps} Representations of the fields under the $A_4 \times Z_4$ and $SU(2)_L \times U(1)_Y$ symmetries.}
\begin{tabular}{ccccccccc}
Field &$L$&$e_R,\mu_R,\tau_R$&$\Delta_1$&$\Delta_2$&$\phi$&$S$&$\Phi$&$\Psi$\\
\hline\hline
$A_4$&$\mathbf{3}$&$\mathbf{1}$, $\mathbf{1^{\prime}}$,
$\mathbf{1^{\prime\prime}}$&$\mathbf{1}$&$\mathbf{1}$&$\mathbf{1}$&$\mathbf{1}$&
$\mathbf{3}$&$\mathbf{3}$\\
$Z_4$&$i$&$-i$&$1$&$-1$&$i$&$-1$&$i$&$1$\\
$SU(2)_L\times U(1)_Y$&$(2,-1/2)$&$(1,-1)$&$(3,1)$&$(3,1)$&$(2,1/2)$&$(1,0)$&$(1,0)$&$(1,0)$\\
\hline
\end{tabular}
\end{table}
\end{center}

Below the cut-off scale $\Lambda$, the flavour dynamics is encoded in the relevant effective Yukawa Lagrangian $\mathcal{L}$, which contains the lowest-order terms\footnote{In principle, one could also include the renormalizable 4-dimension term $\Delta_2 L^T L$. This term is however easily removed by imposing an additional shaping $Z_4$ symmetry.} in an expansion in powers of $1/\Lambda$,
\begin{align}
\begin{split}
\mathcal{L}=&\frac{y^\ell_e}{\Lambda} \left(\overline{L} \Phi\right)_{\mathbf{1}}\phi e_R+\frac{y^\ell_\mu}{\Lambda}  \left(\overline{L} \Phi\right)_{\mathbf{1^{\prime\prime}}}\phi\mu_R+\frac{y^\ell_\tau}{\Lambda}  \left(\overline{L} \Phi\right)_{\mathbf{1^{\prime}}}\phi\tau_R\\
&+\frac{y_{2}}{\Lambda}\, \Delta_2\left(L^TL\Psi\right)_{\mathbf{1}}+\frac{1}{\Lambda} \,\Delta_1\left(L^TL\right)_{\mathbf{1}}\left(y_{1} S+y^{\prime}_{1} S^\ast\right)+\text{H.c.}\,.
\end{split}
\label{LYukA4}
\end{align}
As soon as the heavy scalar fields develop VEVs along the required directions~\footnote{Following the standard procedure, we will assume the typical vacuum alignment for this class of models. The way how this vacuum configuration is achieved from the minimization of the complete flavon potential is out of the scope of the present work.}, namely,
\begin{align}
\langle\Phi\rangle=(r,0,0)\;,\;\langle\Psi\rangle=(s,s,s)\,,
\label{vevalign}
\end{align}
and the scalar singlet $S$ acquires a complex VEV, $\langle S \rangle = v_S\,e^{i\alpha}$, the leptonic mass Lagrangian becomes
\begin{equation}
\begin{aligned}
\label{LYuk}
-\mathcal{L}=&\frac{y^\ell_e\, r}{\Lambda}\,\overline{L}_e\,\phi\, e_R+\frac{y^\ell_\mu\, r}{\Lambda} \overline{L}_\mu\,\phi\,\mu_R+\frac{y^\ell_\tau\, r}{\Lambda}\,\overline{L}_\tau\,\phi\,\tau_R+\frac{y_{2}s}{3\Lambda}\, \Delta_2\left(2L_e^TL_e+2L_\mu^TL_\mu\right.\\
&\left.+2L_\tau^TL_\tau-L_e^TL_\mu-L_e^TL_\tau-L_\mu^TL_e-L_\mu^TL_\tau-L_\tau^TL_e-L_\tau^TL_\mu\right)\\
&+\frac{v_S}{\Lambda^\prime}\,\Delta_1\left(L_e^TL_e + L_\mu^TL_\tau+L_\tau^TL_\mu\right)\left(y_{1} e^{i\alpha} +y^{\prime}_{1} e^{-i\alpha}\right)+\text{H.c.}\\
&\equiv \mathbf{Y}_{\alpha\beta}^e\,\overline{L}_\alpha\,\phi\, e_{R\beta}+\mathbf{Y}_{\alpha\beta}^{\Delta_1}\, L_\alpha^T C \Delta_1 L_\beta+\mathbf{Y}_{\alpha\beta}^{\Delta_2}\, L_\alpha^T C \Delta_2 L_\beta+\text{H.c.},
\end{aligned}
\end{equation}
with
\begin{align}\label{Ymatrices}
\mathbf{Y}^e=
\begin{pmatrix}
y_e&0&0\\
0&y_\mu&0\\
0&0&y_\tau
\end{pmatrix}\,,\quad \mathbf{Y}^{\Delta_1}=y_{\Delta_1}
\begin{pmatrix}
1&0&0\\
0&0&1\\
0&1&0
\end{pmatrix}\,,\quad  \mathbf{Y}^{\Delta_2}=\frac{y_{\Delta_2}}{3}
\begin{pmatrix}
2&-1&-1\\
-1&2&-1\\
-1&-1&2
\end{pmatrix},
\end{align}
and
\begin{align}
y_{e, \mu, \tau} = \frac{r}{\Lambda} y^\ell_{e, \mu, \tau}\,, \quad  y_{\Delta_1}=\frac{v_S}{\Lambda^\prime}\left(y_{1} e^{i\alpha}+y^{\prime}_{1} e^{-i\alpha}\right), \quad y_{\Delta_2}=\frac{y_{2}}{\Lambda}\,s.
\end{align}
Notice that the Yukawa matrices $\mathbf{Y}^{\Delta_1}$ and $\mathbf{Y}^{\Delta_2}$ exhibit the so-called $\mu-\tau$ and magic symmetries, respectively.

\section{Spontaneous CP violation}

In the present model, CP is conserved at the Lagrangian level because all the parameters are set to be real. Yet, this symmetry can be spontaneously broken by the complex VEV of the scalar singlet $S$. To show that this is indeed the case, let us analyze the scalar potential for $S$.  Assuming that this field is very heavy and decouples from the theory at an energy scale much higher than the electroweak and Higgs-triplet scales, the relevant terms in the scalar potential are simply those given by the contribution $V_S$ in Eq.~(\ref{potterms}). The tree-level potential then reads
\begin{align}\label{potV0}
 V_0=m_S^2 v_S^2 + \lambda_S v_S^4 + 2 \left(\mu_S^2+\lambda_S^{\prime\prime} v_S^2\right) v_S^2\,\cos{(2\alpha)}+2 \lambda_S^\prime v_S^4\,\cos{(4\alpha)}.
\end{align}
Its minimization with respect to $v_S$ and $\alpha$ yields the equations
\begin{align}
\frac{\partial V_0}{\partial v_S} &= 2 v_S \left[m_S^2 +2 \lambda_S v_S^2 + 2 (\mu_S^2+2\lambda_S^{\prime\prime} v_S^2)\,\cos{(2\alpha)}+4\lambda_S^\prime v_S^2\,\cos{(4\alpha)}\right] =0,\\
\frac{\partial V_0}{\partial \alpha} &= -4 v_s^2\,\sin{(2\alpha)}\,\left[ \left(\mu_S^2+\lambda_S^{\prime\prime} v_S^2\right)\, + 4\lambda_S^\prime v_S^2\,\cos{(2\alpha)}\right]=0.
\end{align}

Besides the trivial solution $v_S =0$, which leads to $V_0=0$, there are other three possible solutions to the above system of equations:
\begin{align}\label{vssol}
\begin{split}
\text{(i)}&\quad v_S^2=- \frac{m_S^2+2\mu_S^2}{2(\lambda_S+2\lambda_S^\prime+2\lambda_S^{\prime\prime})}\,,\quad \alpha=0, \pm\pi;\\
\text{(ii)} &\quad v_S^2=\frac{-m_S^2+2\mu_S^2}{2(\lambda_S+2\lambda_S^\prime-2\lambda_S^{\prime\prime})}\,,\quad \alpha=\pm\frac{\pi}{2};\\
\text{(iii)} &\quad v_S^2=\frac{-2\lambda_S^\prime m_S^2+\lambda_S^{\prime\prime}\mu_S^2}{4\lambda_S\lambda_S^\prime - 8\lambda_S^{\prime 2} - \lambda_S^{\prime\prime 2}}\,,\quad \cos(2 \alpha) = -\frac{\mu_S^2+\lambda_S^{\prime\prime}v_S^2}{4\lambda_S^\prime v_S^2}\,.
\end{split}
\end{align}
Note that in spite of the phase $\pi/2$ in (ii), in this case the vacuum does not violate CP~\cite{Branco:1999fs}. Therefore, only the last solution is of interest to us since it leads not only to the spontaneous breaking of the CP symmetry but also to a non-trivial CP-violating phase in the one-loop diagrams relevant for leptogenesis. One can show that this solution indeed corresponds to the global minimum of the potential in a wide region of the parameter space. To illustrate this, let us consider  $m_S^2<0, \lambda_S^{\prime\prime}\simeq 0, \mu_S \simeq 0$ and $\lambda_S>2\lambda_S^{\prime}>0$. From Eqs.~(\ref{vssol}) we then obtain
\begin{align}
v_S^2 \simeq-\frac{m_S^2}{2(\lambda_S+2\lambda_S^\prime)}\,,\quad \alpha=0, \pm\frac{\pi}{2}, \pm\pi,
\label{iandii}
\end{align}
for the cases (i) and (ii), leading to
\begin{align}
V_0 \simeq -\frac{m_S^4}{4\left(\lambda_S+2\lambda^\prime_S\right)}.
\end{align}
In turn, solution (iii) yields
\begin{align}
v_S^2\simeq-\frac{m_S^2}{2(\lambda_S-2\lambda_S^{\prime })}\,,\quad \alpha \simeq \pm \pi/4\,,
\end{align}
implying
\begin{align}
V_0 \simeq -\frac{m_S^4}{4\left(\lambda_S-2\lambda^\prime_S\right)}.
\end{align}
Clearly, the latter value corresponds to the absolute minimum of the potential. One can also easily show that both mass eigenvalues are positive within the assumed parameter region: $M_{S,1}^2=-4 m^2_S>0$ and $M_{S,2}^2=-16m_S^2\lambda_S^\prime/(\lambda_S-2\lambda_S^\prime)>0$.

\section{Seesaw mechanism, neutrino masses and leptonic mixing}
\label{sawneut}

In the present framework, neutrinos acquire masses through the well-known type II seesaw mechanism due to the tree-level exchange of the heavy scalar triplets $\Delta_a$. From Eq.~(\ref{LYuk}) it is straightforward to see that the effective neutrino mass matrix $\mathbf{m}_\nu$ is given by
\begin{align} \label{uamu}
\mathbf{m}_\nu=\mathbf{m}_\nu^{(1)}+\mathbf{m}_\nu^{(2)}, \quad \mathbf{m}_\nu^{(a)}=2 u_a \mathbf{Y}^{\Delta_a},
\end{align}
where $u_a = \mu_a^*\, v^2/M_a$ are the VEVs of the neutral components $\Delta^0_a$ of the scalar triplets, $\langle \phi_0 \rangle = v=174$~GeV and, in accordance with Eq.~(\ref{potterms}), $\mu_{1}=\left(\lambda_{1} e^{i\alpha}+\lambda^{\prime}_{1} e^{-i\alpha}\right)v_S/M_1$. The Yukawa matrices $\mathbf{Y}^{\Delta_a}$ are those already presented in Eq.~(\ref{Ymatrices}).
Diagonalizing $\mathbf{m}_\nu$ we obtain
\begin{align} \label{Mnudiag}
\mathbf{m}_\nu=\mathbf{U}^\ast\, \mathbf{d}_\nu\, \mathbf{U}^\dagger\,,\quad \mathbf{d}_\nu=\text{diag}\left(|z_1 e^{i\beta}+z_2|,z_1,|z_1 e^{i\beta} - z_2|\right) \equiv \text{diag}(m_1,m_2,m_3),
\end{align}
where $m_i$ are the neutrino masses and
\begin{align}\label{zaua}
z_a=2 |u_a y_{\Delta_a}|\,,\quad \beta=\text{arg}(u_1 y_{\Delta_1})=
\arctan\left(\frac{\lambda_1^\prime-\lambda_1}{\lambda_1+\lambda_1^\prime}\tan\alpha\right)+ \arctan\left(\frac{y_1-y_1^\prime}{y_1+y_1^\prime}\tan\alpha\right)\,.
\end{align}
Hereafter we consider the relevant CP-violating phase as being $\beta$. The unitary matrix $\mathbf{U}$ is given by
\begin{align}
\mathbf{U}= e^{-i\sigma_1/2}\, \mathbf{U}_\text{TBM}\, \mathbf{K},
\end{align}
where $\mathbf{U}_\text{TBM}$ is the tribimaximal mixing matrix,
\begin{align} \label{utbm}
\mathbf{U}_\text{TBM}=\begin{pmatrix}
\frac{2}{\sqrt{6}}&\frac{1}{\sqrt{3}}&0\\
-\frac{1}{\sqrt{6}}&\frac{1}{\sqrt{3}}&-\frac{1}{\sqrt{2}}\\
-\frac{1}{\sqrt{6}}&\frac{1}{\sqrt{3}}&\frac{1}{\sqrt{2}}
\end{pmatrix},
\end{align}
and
\begin{align}
\begin{split}
\mathbf{K} &=\text{diag}\left(1,e^{i\gamma_1},e^{i\gamma_2}\right),\\
\gamma_1 &= (\sigma_1-\beta)/2, \quad \gamma_2=(\sigma_1-\sigma_2)/2, \quad \sigma_{1,2}=\text{arg}\left(z_2 \pm z_1 e^{i\beta}\right).
\end{split}
\end{align}
Since at this point there is no Dirac-type CP violation ($\mathbf{U}_{13}=0$), the Majorana phases $\gamma_{1,2}$ are the only source of CP violation in the lepton sector.

Let us now discuss how the experimental knowledge on the neutrino mass squared differences constrains the parameters $z_1$ and $z_2$. At $1\sigma$ confidence level, the neutrino mass squared differences are~\cite{Schwetz:2011zk}
\begin{align}\label{expdmass}
\Delta m^2_{21} =\left(7.59^{+0.20}_{-0.18}\right)\times 10^{-5}\,\text{eV}^2,\quad \Delta m^2_{31}=\left(2.50^{+0.09}_{-0.16}\right)\left[-2.40^{+0.09}_{-0.08}\right]
\times 10^{-3}\,\text{eV}^2,
\end{align}
for the normal [inverted] neutrino mass hierarchy. The sign of the neutrino mass difference $(m_3 - m_2)$ is dictated by the ordering of the neutrino masses: positive for normal ordering ($m_3>m_2>m_1$) and negative for inverted ordering ($m_3<m_2<m_1$). In the first case, Eq.~(\ref{Mnudiag}) together with the condition $m_3> m_2 > m_1$ implies the constraints
\begin{align}
z_2+2z_1\cos\beta<0, \quad z_2-2z_1\cos\beta>0,
\end{align}
which, clearly, are satisfied only if $\pi/2 < \beta < 3\pi/2$. On the other hand, for $m_3<m_2<m_1$ one has
\begin{align}
z_2+2z_1\cos\beta<0, \quad z_2-2z_1\cos\beta<0,
\end{align}
which cannot be simultaneously fulfilled. Thus, the present model cannot accommodate an inverted hierarchy for the neutrino mass spectrum.

The parameters $z_1$ and $z_2$ can be written in terms of $\Delta m_{21}^2$, $\Delta m_{31}^2$ and the angle $\beta$ as
\begin{align}\label{z1z2}
\begin{split}
z_1&=-\frac{1}{2\cos\beta}\frac{\Delta m_{31}^2}{\sqrt{2(\Delta m_{31}^2-2\Delta m_{21}^2)}} \simeq -\frac{1}{2\cos\beta}\sqrt{\frac{\Delta m_{31}^2}{2}}\,,\\
z_2&=\sqrt{\frac{\Delta m_{31}^2-2\Delta m_{21}^2}{2}} \simeq \sqrt{\frac{\Delta m_{31}^2}{2}},
\end{split}
\end{align}
where the last equalities in the right-hand sides were obtained using the fact that $\Delta m_{31}^2 \gg \Delta m_{21}^2$. We notice that $z_2$ is completely fixed by the atmospheric neutrino mass squared difference. Moreover, $z_1 \simeq -z_2/(2 \cos \beta)$. Using the best fit values given in Eqs.~(\ref{expdmass}), we obtain
\begin{align}\label{z1z2best}
z_1 \simeq -0.0175/\cos \beta \ge 0.0175~\text{eV}, \quad z_2 \simeq 0.035~\text{eV}.
\end{align}
In turn, the neutrino masses defined in Eq.~(\ref{Mnudiag}) can be rewritten as
\begin{align}
m_1=\sqrt{z_1^2-\Delta m_{21}^2}\,,\quad m_2=z_1,\quad m_3=\sqrt{z_1^2+\Delta m_{31}^2-\Delta m_{21}^2}\,.
\end{align}
Thus, the model predicts a lower bound for the lightest neutrino mass: $m_1 \gtrsim 1.5\times 10^{-2}$~eV. The neutrino mass hierarchy is maximal when $\beta=\pi$, while an almost degenerate spectrum is obtained for $\beta \simeq \pi/2$ or $\beta \simeq 3\pi/2$. Finally, the Majorana phases are approximately given by
\begin{align}
\gamma_1 \simeq -\beta, \quad \gamma_2 \simeq - \frac{\beta}{2} - \frac12 \arctan \left(\frac{\tan\beta}{3}\right).
\end{align}

\begin{figure}[t]
\begin{center}
\includegraphics[width=10cm]{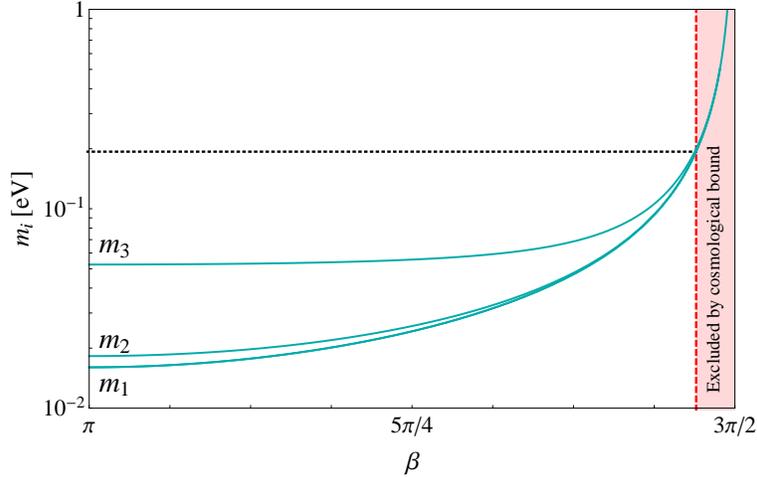}
\caption{\label{fig1} Neutrino masses $m_i$ as a function of the high-energy phase $\beta$ in the exact TBM case.}
\end{center}
\end{figure}

The dependence of neutrino masses on the high-energy phase $\beta$ is presented in Fig.~\ref{fig1} for the exact TBM case. The light red shaded area is currently disfavoured by the recent WMAP seven-year cosmological observational data~\cite{Komatsu:2010fb}. Although WMAP alone constrains the sum of light neutrino masses below 1.3 eV (95\% CL), when combined with baryonic acoustic oscillation and type-Ia supernova data this bound is more restrictive: $\sum_i\,m_i<0.58$~eV at 95\% CL.

\subsection{Vacuum-alignment perturbations and nonzero $\theta_{13}$}

The T2K~\cite{Abe:2011sj} and MINOS~\cite{Adamson:2011qu} neutrino oscillation data imply for the $\theta_{13}$ mixing angle
\begin{align}
 \sin^2\theta_{13}=0.013^{+0.007}_{-0.005}\left(^{+0.015}_{-0.009}\right)\left[^{+0.022}_{-0.012}\right]\,,
\end{align}
at $1\sigma(2\sigma)[3\sigma]$. Recently, through the observation of electron-antineutrino disappearance, the Daya Bay Reactor Neutrino Experiment has also measured the non-zero value~\cite{DayaBay}:
\begin{equation}
\sin^2(2\theta_{13})=0.092\pm0.016({\rm stat})\pm 0.005({\rm syst})\,,
\label{DBresults}
\end{equation}
with a significance of $5.2\sigma$. In the light of these results, models that lead to tribimaximal mixing appear to be disfavored. In general, deviations from $\theta_{13}=0$ in these models cannot bring this angle into agreement with data without spoiling the predictions for the solar and atmospheric mixing angles. Several alternative solutions have been recently put forward to explain a relatively large value of $\theta_{13}$ in the framework of discrete flavour symmetries (see, for instance, Refs.~\cite{Branco:2009by,Ahn:2011yj,Ma:2011yi,Meloni:2011fx,Morisi:2011pm,Toorop:2011jn,Ge:2011qn,Wu:2012ri,Meloni:2012sy}).

Here we shall follow an approach based on considering small perturbations around the TBM vacuum-alignment conditions~(\ref{vevalign}). Such perturbations could come from the presence of higher dimensional operators in the flavon potential. We consider two distinct cases:

{\bf CASE A} - Small perturbations around the flavon VEV $\left<\Phi\right>=(r,0,0)$ of the form $\left<\Phi\right>=r(1,\varepsilon_1,\varepsilon_2)$;\\

{\bf CASE B} - Small perturbations around the flavon VEV $\left<\Psi\right>=s(1,1,1)$ of the form $\left<\Psi\right>=s(1,1+\varepsilon_1,1+\varepsilon_2)$;

\noindent with $|\varepsilon_{1,2}|\ll 1$.

Since $\langle \Psi \rangle$ is not perturbed in case A, the contribution to the lepton mixing coming from the neutrino sector remains of TBM type [cf. Eq.~(\ref{LYukA4})]. On the other hand, due to the new form of $\left<\Phi\right>$, the charged lepton Yukawa matrix is
\begin{equation}
\mathbf{Y}^\ell=
\begin{pmatrix}
y_e&y_\tau\varepsilon_1&y_\mu\varepsilon_2\\
y_\tau \varepsilon_2&y_\mu&y_e\varepsilon_1\\
y_\mu\varepsilon_1&y_e\varepsilon_2&y_\tau
\end{pmatrix}\,,
\end{equation}
which implies $\mathbf{U}_\ell\neq 1\!\!1$, where $\mathbf{U}_\ell$ is the unitary matrix which rotates the left-handed charged-lepton fields to the their physical basis. The new lepton mixing matrix $\mathbf{U}=\mathbf{U}_\ell^\dagger \mathbf{U}_{\rm TBM}$ yields  the perturbed mixing angles
\begin{equation}
\sin^2\theta_{12}\simeq\frac{1}{3}\left[1-2(\varepsilon_1+\varepsilon_2)\right]\,,\quad \sin^2\theta_{23}\simeq\frac{1}{2}(1+2\varepsilon_1)\,,\quad
\sin^2\theta_{13}\simeq \frac{\left(\varepsilon_1-\varepsilon_2\right)^2}{2}\,,
\label{aps}
\end{equation}
at lowest order in $\varepsilon_{1,2}$. This leads to the following approximate relation among the three lepton mixing angles
\begin{equation}
\sin^2\theta_{13}\simeq \frac{(4\sin^2\theta_{23}-3\cos^2\theta_{12})^2}{8}\,,
\label{aprel}
\end{equation}
compatible with neutrino data at the $1\sigma$ level. Obviously, the rotation of the charged lepton fields does not affect the neutrino spectrum nor generate a Dirac-type CP-violating phase. Since the flavon fields are real, the Majorana phases $\gamma_{1,2}$ also remain unaltered.

\begin{figure}[t]
\begin{tabular}{cc}
\includegraphics[width=7.0cm]{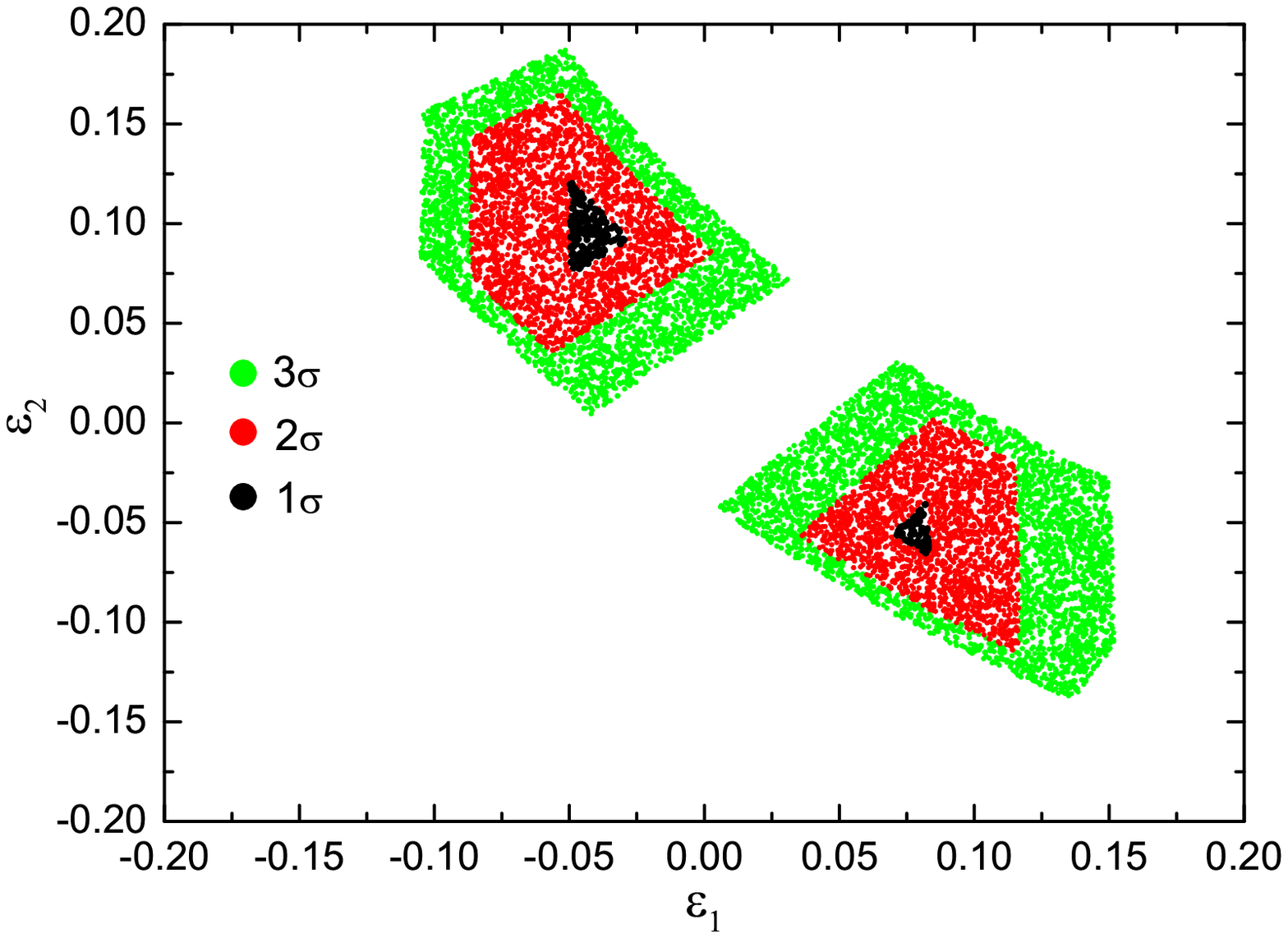}&
\includegraphics[width=7.0cm]{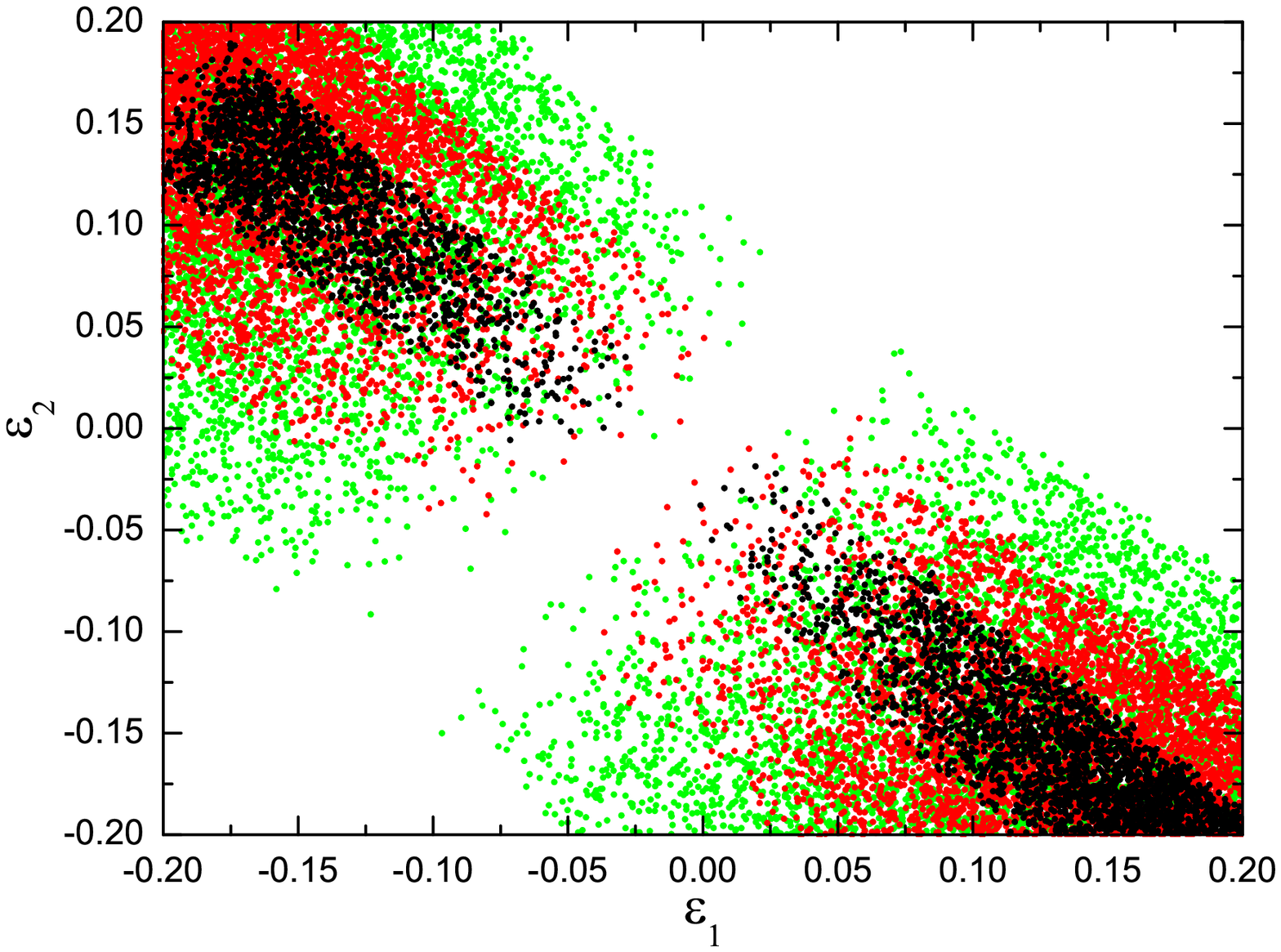}
\end{tabular}
\caption{Allowed regions in the $(\varepsilon_1,\varepsilon_2)$ plane corresponding to the VEV perturbations of the flavon field $\left<\Phi\right>=r(1,\varepsilon_1,\varepsilon_2)$ in case A (left panel) and $\left<\Psi\right>=s(1,1+\varepsilon_1,1+\varepsilon_2)$ in case B (right panel). The scatter points were obtained considering the $1\sigma$ (black), $2\sigma$ (red) and
$3\sigma$ (green) neutrino oscillation data.\label{fig2}}
\end{figure}

In Fig.~\ref{fig2} (left panel) we present the allowed regions in the $(\varepsilon_1,\varepsilon_2)$ plane taking into account the present neutrino oscillation data, at $1\sigma$ (black), $2\sigma$ (red) and
$3\sigma$ (green). As it is apparent from the figure, with $(\varepsilon_1,\varepsilon_2)\simeq  (-0.05, 0.1)$ or $(\varepsilon_1,\varepsilon_2)\simeq  (0.05, -0.05)$ one obtains agreement with the data at the $1\sigma$ level.

\begin{figure}[t]
\centering
\includegraphics[width=10.0cm]{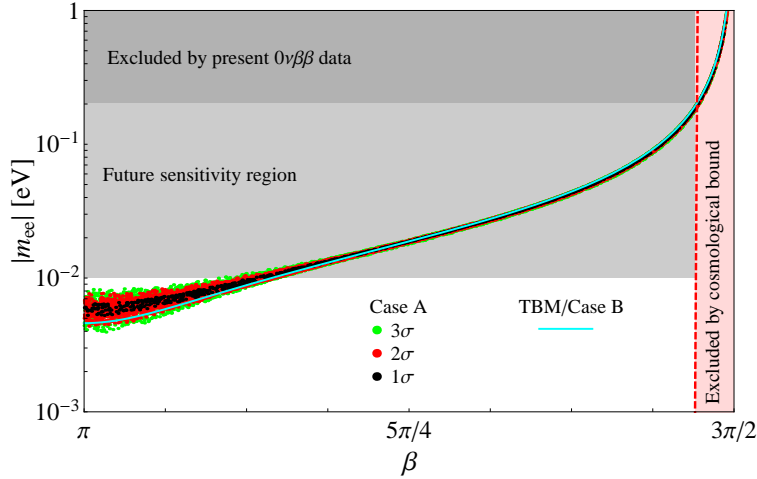}
\caption{\label{fig3} Neutrinoless double beta decay parameter $|m_{ee}|$ as a function of the high-energy phase $\beta$ for case A (scatter points) and TBM and case B (cyan solid line).}
\end{figure}

Neutrinoless double beta decay ($0\nu\beta\beta$) is an important low-energy process~\cite{Rodejohann:2011mu} which, if observed, will establish the Majorana nature of neutrinos. The rate of this process is proportional to the modulus of the (11) entry of the effective neutrino mass matrix, in the weak basis where the charged-lepton mass matrix is diagonal and real. In our framework its value is given by
\begin{align}\label{mee1}
|m_{ee}| = \left| \sum_i m_i\,\mathbf{U}_{1i}^{2}\right|= \frac{1}{3}\,\left| 2\,m_1(1+\varepsilon_1+\varepsilon_2) +m_2\,e^{-2 i\gamma_1}\right|\,,
\end{align}
in leading order of $\varepsilon_{1,2}$.
Although with large uncertainties from the poorly known nuclear matrix elements, data available at present set an upper bound on $|m_{ee}|$ in the range 0.2 to 1 eV at 90\% C.L.~\cite{KlapdorKleingrothaus:2000sn,Arnaboldi:2008ds,Wolf:2008hf}. The existing limits will be considerably improved in the forthcoming experiments, with an expected sensitivity of about $10^{-2}$~eV~\cite{Aalseth:2004hb}. In Fig.~\ref{fig3} we present the dependence of $|m_{ee}|$ as a function of the phase $\beta$ for case A (scatter points) taking into account the present neutrino oscillation data, at $1\sigma$ (black), $2\sigma$ (red) and $3\sigma$ (green). We obtain $|m_{ee}| \simeq (0.004-0.2)$~eV, where the upper limit comes from the cosmological bound and it corresponds to an almost degenerate neutrino spectrum. We notice that the predictions for the effective Majorana mass parameter $|m_{ee}|$ are within the reach of future experiments~\cite{Rodejohann:2011mu} and the model can be ruled out if $0\nu\beta\beta$ searches give $|m_{ee}| < 4$~meV.

We now discuss case B, in which $\left<\Phi\right>=(r,0,0)$ and small perturbations around the flavon VEV $\left<\Psi\right>=s(1,1,1)$ of the type $\left<\Psi\right>=s(1,1+\varepsilon_1,1+\varepsilon_2)$ are considered. This will induce corrections to the mixing coming from the neutrino sector. Notice that one can alternatively consider perturbations of the type $\left<\Psi\right>=s^\prime(1+\varepsilon_1^\prime,1,1+\varepsilon_2^\prime)$ or $\left<\Psi\right>=s^{\prime\prime}(1+\varepsilon_1^{\prime\prime},1+\varepsilon_2^{\prime\prime},1)$. The results obtained below are obviously invariant under the choice of the perturbation, provided one takes into account the mapping of $\{s^\prime, s^{\prime\prime},\varepsilon_i^\prime,\varepsilon_i^{\prime\prime}\}$ into $\{s,\varepsilon_i\}$. At leading order this mapping corresponds to $s^\prime=s^{\prime\prime}=s(1+\varepsilon_1)$, $\varepsilon_1^\prime=-\varepsilon_1$, $\varepsilon_2^\prime=\varepsilon_2-\varepsilon_1$, $\varepsilon_1^{\prime\prime}=\varepsilon_2-\varepsilon_1$ and $\varepsilon_1^{\prime\prime}=-\varepsilon_1$.

The Yukawa couplings $\mathbf{Y}^{\Delta_{2}}$ contributing to the neutrino mass matrix are now given by
\begin{equation}
\mathbf{Y}^{\Delta_{2}}=\frac{y_{\Delta_2}}{3}
\begin{pmatrix}
2&&-1-\varepsilon_2&&-1-\varepsilon_1\\
-1-\varepsilon_2&&2+2\varepsilon_1&&-1\\
-1-\varepsilon_1&&-1&&2+2\varepsilon_2
\end{pmatrix}.
\end{equation}
Consequently, at first order in $\varepsilon_{1,2}$, the neutrino mass spectrum gets corrected in the following way
\begin{align}
\begin{split}
m_1^2\simeq&\frac{1}{3}\left[3z_1^2+z_2^2\left(3+2\varepsilon_1+2\varepsilon_2\right)
+z_1z_2(3+\varepsilon_1+\varepsilon_2)\cos\beta\right],\\
m_2^2\simeq&z_1^2,\\
m_3^2\simeq&\frac{1}{3}\left[3z_1^2+z_2^2\left(3+2\varepsilon_1+2\varepsilon_2\right)
-z_1z_2(3+\varepsilon_1+\varepsilon_2)\cos\beta\right]\,.
\end{split}
\end{align}
As in the unperturbed case, it can be shown from the above equations that an inverted neutrino hierarchy is not allowed.

In the present case, the approximate analytic expressions for the mixing angles are
\begin{equation}
\sin^2\theta_{12}\simeq\frac{1}{3}+\frac{2}{9}(\varepsilon_1+\varepsilon_2),\quad  \sin^2\theta_{13}\simeq\frac{(\varepsilon_1-\varepsilon_2)^2}{72\cos^2\beta},\quad
\sin^2\theta_{23}\simeq\frac{1}{2}+\frac{1}{6}(\varepsilon_1-\varepsilon_2)\,,
\label{apsB}
\end{equation}
while for the Dirac-type CP-violating invariant $J_{\rm CP}$ we have
\begin{equation}
J_{\rm CP}={\rm Im}\left[\,\mathbf{U}_{11} \mathbf{U}_{22}
\mathbf{U}_{12}^\ast \mathbf{U}_{21}^\ast\,\right] \nonumber \\
\simeq \frac{\varepsilon_2-\varepsilon_1}{36}\tan\beta\,.
\label{JCP}
\end{equation}
Using the standard form $J_{\rm CP}=  \sin(2\,\theta_{12})
\sin(2\,\theta_{13})\sin(2\,\theta_{23}) \sin \delta/8$ together with  the relations (\ref{apsB}), we get $\sin\delta\simeq\sin\beta$, which means that the Dirac CP-violating phase $\delta$ is directly related with the phase of the singlet VEV $\langle S \rangle$, as shown in Eq.~(\ref{zaua}). The numerical results for the allowed regions in the $(\varepsilon_1,\varepsilon_2)$ plane in case B are shown in Fig.~\ref{fig2} (right panel). By comparing both panels in the figure, it is clear that case B is less restrictive than case A. We also note that the predictions for $0\nu\beta\beta$ are exactly the same as for the TBM case (cyan solid line in Fig.~\ref{fig3}) so that  $m_{ee}$ is obtained from Eq.~(\ref{mee1}) in the limit $\varepsilon_{1,2}=0$.

\begin{figure}[t]
\begin{tabular}{cc}
\includegraphics[width=7.0cm]{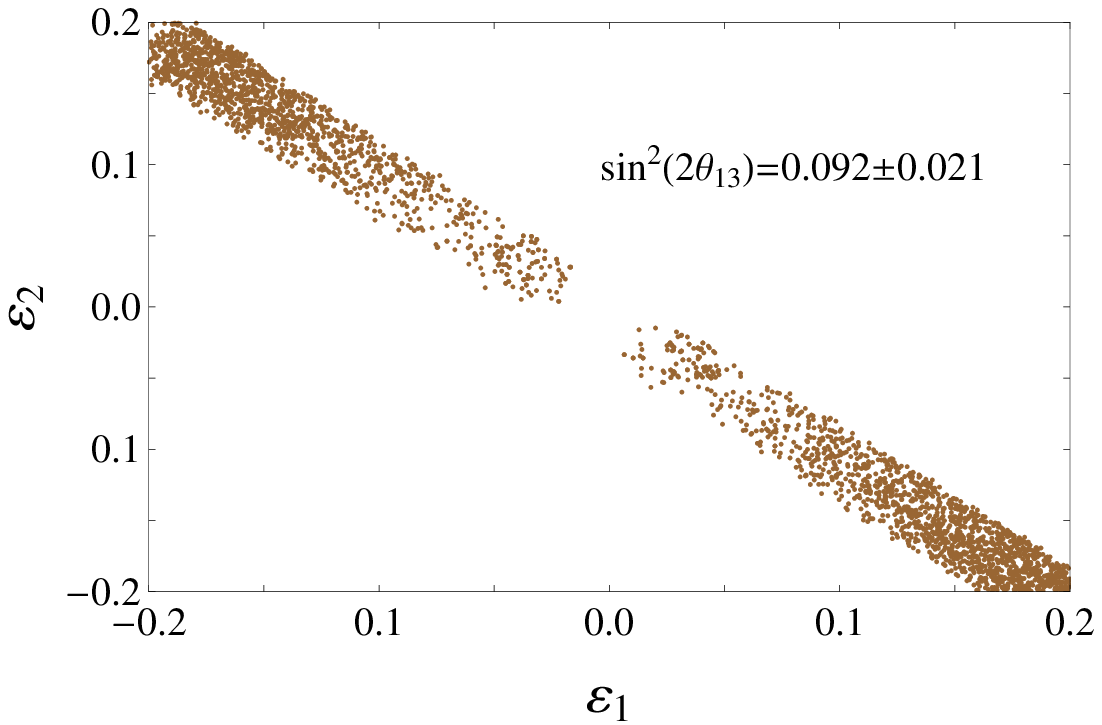}&
\includegraphics[width=7.0cm]{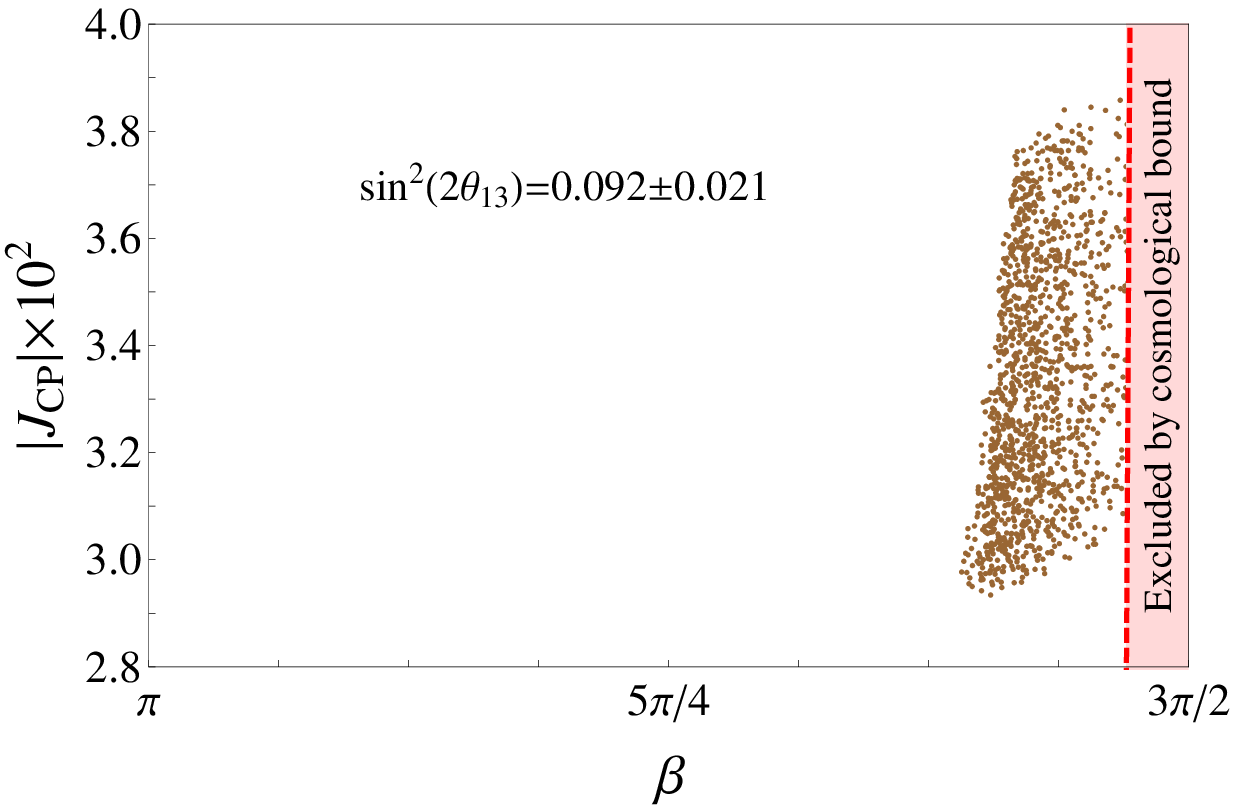}
\end{tabular}
\caption{Left: allowed regions in the $(\varepsilon_1,\varepsilon_2)$ plane corresponding to the VEV perturbations of the flavon field $\Psi$ in case B, taking into account the Daya Bay result for $\theta_{13}$. The corresponding regions in the $(|J_{\rm CP}|,\beta)$ plane are shown in the right panel.}
\label{fig4a}
\end{figure}

We now comment on the possibility of reproducing the recent Daya Bay $\theta_{13}$ value (\ref{DBresults}) in our framework. In the absence of a 3-neutrino global analysis of the oscillation data including the Daya Bay results, we take the $1\sigma$ values for $\theta_{12}$, $\theta_{23}$ and $\Delta m^2_{21,31}$ obtained in \cite{Schwetz:2011zk}. From Eq.~(\ref{aprel}) one can see that the new Daya Bay value for $\theta_{13}$ is not compatible with the remaining mixing angles for case A. Instead, for case B we get a perfect agreement with all data. This is apparent from Fig.~\ref{fig4a} (left panel)  where the allowed regions in the $(\varepsilon_1,\varepsilon_2)$ plane are shown. In the right panel of the same figure the predictions for $|J_{\rm CP}|$ are shown as a function of the CP-violating phase $\beta$ which, as already mentioned, is approximately equal to the Dirac phase in the lepton mixing matrix. From these results we conclude that our framework predicts $11\pi/8 \lesssim \delta\lesssim 3\pi/2$ with $0.03\lesssim |J_{\rm CP}|\lesssim 0.04$, which is large enough to be measured in future oscillation experiments.

\section{Higgs triplet decays and leptogenesis}

\begin{figure}[t]
\centering
\includegraphics{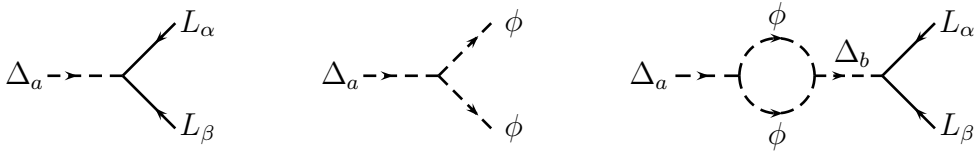}\\
\caption{\label{fig4} Tree-level diagrams for the scalar triplet decays and one-loop diagram contributing to the CP asymmetry $\epsilon_{a}^{\alpha\beta}$.}
\end{figure}

The mechanism of leptogenesis can be naturally realized in the present model due to the presence of the scalar triplets $\Delta_1$ and $\Delta_2$. At tree level, the latter can decay into two leptons or two Higgs fields (cf. Fig.~\ref{fig4}).
In the presence of CP-violating interactions, the decay of $\Delta_a$ into two leptons generates a nonvanishing leptonic asymmetry for each triplet component ($\Delta^0_a, \Delta^+_a, \Delta^{++}_a$),
\begin{align} \label{cpasymII}
\epsilon_{a} = 2 \sum_{\alpha\beta} {{\Gamma (\Delta_a^* \rightarrow L_\alpha + L_\beta) - \Gamma (\Delta_a \rightarrow \bar{L}_\alpha + \bar{L}_\beta)}\over{\Gamma_{\Delta_a} +\Gamma_{\Delta^*_a}}}\,,
\end{align}
where $\Gamma_{\Delta_a}$ denotes the total triplet decay width and the overall factor of 2 arises because the triplet decay produces two leptons. It is useful to define
\begin{align}\label{BlBphi}
    \mathcal{B}^L_a\, \Gamma_{\Delta_a} &\equiv \sum_{\alpha,\beta} \Gamma(\Delta_a^* \rightarrow L_\alpha+L_\beta)=\frac{M_a}{8\pi} \text{Tr}\, (\mathbf{Y}^{\Delta_a\dagger} \mathbf{Y}^{\Delta_a}),\nonumber\\
    \mathcal{B}^\phi_a \Gamma_{\Delta_a} &\equiv \Gamma(\Delta_a^* \rightarrow \phi+\phi)
    =\frac{M_a}{8\pi}|\mu_a|^2\,,
\end{align}
where $\mathcal{B}^L_a$ and $\mathcal{B}^\phi_a$ are the tree-level branching ratios to leptons and Higgs doublets, respectively. The total triplet decay width is given by
\begin{align}\label{tripletdecaywidth}
    \Gamma_{\Delta_a} = \frac{M_a}{8\pi} \left[ \text{Tr}(\, \mathbf{Y}^{\Delta_a\dagger} \mathbf{Y}^{\Delta_a}) + |\mu_a|^2\right].
\end{align}

When the triplet decays into leptons with given flavours $L_\alpha$ and $L_\beta$, a nonvanishing asymmetry $\epsilon_{a}^{\alpha\beta}$ is generated by the interference of the tree-level decay process with the one-loop self-energy diagram shown in Fig.~\ref{fig4}. One finds~\cite{Branco:2011zb}
\begin{align}\label{cpasymIIa}
\epsilon_{a}^{\alpha\beta} \simeq -\frac{g(x_b)}{2\pi} \frac{c_{\alpha\beta}\,\text{Im}\bigl[\mu_a^\ast\mu_b \mathbf{Y}^{\Delta_a}_{\alpha\beta} \mathbf{Y}^{\Delta_b\ast}_{\alpha\beta} \bigr]}{\text{Tr}\left(\mathbf{Y}^{\Delta_a\dagger} \mathbf{Y}^{\Delta_a}\right)+\left|\mu_a\right|^2}, \quad (b \neq a),
\end{align}
where
\begin{equation}\label{cab}
c_{\alpha\beta}=\left\{\begin{array}{ll} 2-\delta_{\alpha\beta} &\quad \text{for}\quad \Delta^0_a, \Delta^{++}_a\\ 1 &\quad \text{for}\quad \Delta^+_a
\end{array}\right.,
\end{equation}
$x_{b} = M_b^2/M_a^2$, and the one-loop self-energy function $g(x_b)$ reads
\begin{align}\label{loopfunc}
   g(x_b) = \frac{\sqrt{x_b}\,(1-x_b)}{(x_b-1)^2+(\Gamma_{\Delta_b}/M_a)^2}.
\end{align}

Recalling that the effective light neutrino mass matrix is given by Eq.~(\ref{uamu}) in the type-II seesaw framework under discussion, Eq.~(\ref{cpasymIIa}) can be rewritten as
\begin{align}\label{cpasymIIb}
   \epsilon_{a}^{\alpha\beta} \simeq -\frac{g(x_b)}{4\pi} \frac{M_b(\mathcal{B}^L_a\, \mathcal{B}^\phi_a)^{1/2}}{v^2}\,\frac{c_{\alpha\beta} \mathrm{Im}\bigl[\mathbf{m}_{\nu,\alpha\beta}^{(a)} \mathbf{m}_{\nu,\alpha\beta}^{\ast}\bigr]}{\bigl[\mathrm{Tr}(\mathbf{m}_\nu^{(a)^\dagger} \mathbf{m}_\nu^{(a)})\bigr]^{1/2}}\,.
\end{align}

In the hierarchical limit $M_a \ll M_b$, it reduces to
\begin{align}\label{cpasymIIc}
   \epsilon_{a}^{\alpha\beta} \simeq \frac{M_a(\mathcal{B}^L_a\, \mathcal{B}^\phi_a)^{1/2}}{4\pi v^2}\,\frac{c_{\alpha\beta}\, \mathrm{Im}\bigl[\mathbf{m}_{\nu,\alpha\beta}^{(a)} \mathbf{m}_{\nu,\alpha\beta}^{\ast}\bigr]}{\bigl[\mathrm{Tr}(\mathbf{m}_\nu^{(a)^\dagger} \mathbf{m}_\nu^{(a)})\bigr]^{1/2}}\,,
\end{align}
which, after summing over the final lepton flavours, yields the following expression for the unflavoured asymmetry~\cite{Hambye:2005tk,Dorsner:2005ii}:
\begin{align}\label{cpasymIId}
    \epsilon_{a}= \frac{M_a(\mathcal{B}^L_a\, \mathcal{B}^\phi_a)^{1/2}}{4\pi v^2}\,\frac{\mathrm{Im}\bigl[\mathrm{Tr}( \mathbf{m}_{\nu}^{(a)} \mathbf{m}_{\nu}^{\dagger})\bigr]}{\bigl[\mathrm{Tr}(\mathbf{m}_\nu^{(a)^\dagger} \mathbf{m}_\nu^{(a)})\bigr]^{1/2}}\,.
\end{align}
One can then show that the following upper bound holds~\cite{Hambye:2005tk}:
\begin{align} \label{epsmaxII}
  |\epsilon_{a}| &\leq \frac{M_a (\mathcal{B}^L_a\, \mathcal{B}^\phi_a)^{1/2}}{4\pi v^2}\,
  \bigl[\mathrm{Tr}(\mathbf{m}_\nu^\dagger \mathbf{m}_\nu)\bigr]^{1/2} =\frac{M_a(\mathcal{B}^L_a\, \mathcal{B}^\phi_a)^{1/2}}{4\pi v^2}\,
  \Bigl( \sum_k m_k^2\Bigr)^{1/2}.
\end{align}
This upper bound increases as the light neutrino mass scale increases. For hierarchical light neutrinos one obtains:
\begin{align}\label{epsmaxIIa}
    |\epsilon_a| \lesssim 10^{-6}\bigl(\mathcal{B}^L_a\, \mathcal{B}^\phi_a\bigr)^{1/2} \left(\frac{M_a}{10^{10}\,\text{GeV}}\right)\biggl(\frac{\sqrt{\Delta m^2_{31}}}{0.05\,\text{eV}}\biggr).
\end{align}
Clearly, the absolute maximum of the right-hand sides of Eqs.~(\ref{epsmaxII}) and (\ref{epsmaxIIa}) is obtained when $\mathcal{B}^L_a = \mathcal{B}^\phi_a=1/2$. Nevertheless, since the efficiency of leptogenesis is dictated by the solution of the relevant Boltzmann equations, the final baryon asymmetry is not necessarily maximal in such a case.

Due to the specific flavour structure of the Yukawa coupling matrices $\mathbf{Y}^{\Delta_a}$ given in Eqs.~(\ref{Ymatrices}), the quantity $\text{Tr}\,\bigl(\mathbf{m}_{\nu}^{(a)} \mathbf{m}_{\nu}^{(b)\dagger}\bigr)$ vanishes in the exact TBM case. This conclusion also holds when the flavon VEV perturbations corresponding to cases A and B are considered. Therefore, the leptogenesis asymmetry defined in Eq.~(\ref{cpasymIId}) is equal to zero and unflavoured leptogenesis is suppressed\footnote{
Unlike the type-I seesaw framework~\cite{Bertuzzo:2009im,AristizabalSierra:2009ex,Felipe:2009rr}, imposing to the Lagrangian a discrete symmetry, such as the $A_4$ symmetry, would not necessarily lead to the vanishing of the leptogenesis asymmetry in the type II seesaw case~\cite{deMedeirosVarzielas:2011tp}. Indeed, even if both matrices, $\mathbf{m}_\nu^{(a)}$ and $\mathbf{m}_\nu^{(b)}$, are diagonalized by the tribimaximal mixing matrix $\mathbf{U}_\text{TBM}$, i.e. $\mathbf{m}_\nu^{(a)}=\mathbf{U}_\text{TBM}\,\mathbf{K}_a^\ast\, \mathbf{d}_\nu^{(a)}\,\mathbf{K}_a^\ast\, \mathbf{U}_\text{TBM}^T$, the quantity $\mathrm{Im}\bigl[\text{Tr}(\mathbf{m}_{\nu}^{(a)} \mathbf{m}_{\nu}^{(b)\dagger})\bigr]=\mathrm{Im}\bigl\{\text{Tr}\bigl[(\mathbf{K}_a^\ast \mathbf{K}_b)^2 \mathbf{d}_\nu^{(a)}\mathbf{d}_\nu^{(b)}\bigr]\bigr\}$ in general does not vanish.}. This means that for leptogenesis to be viable in our framework it must take place in the flavoured regime. In particular, for $T \lesssim 10^{12}$~GeV ($T \lesssim 10^9$~GeV) interactions involving the $\tau$  ($\mu$) Yukawa coupling are in thermal equilibrium and the corresponding lepton doublet is a distinguishable mass eigenstate.

Assuming $M_a \ll M_b$ and using Eqs.~(\ref{Ymatrices}), (\ref{uamu}), (\ref{zaua}) and (\ref{BlBphi}), the CP asymmetry given in Eq.~(\ref{cpasymIIc}) for each triplet component can be rewritten as
\begin{align}\label{CPasymm1}
\epsilon_{a}^{\alpha\beta}=c_{\alpha\beta}\,\mathbf{P}_{\alpha\beta}^a\,\epsilon_a^0\,,
\end{align}
where $c_{\alpha\beta}$ is defined in Eq.~(\ref{cab}), and
\begin{equation}\label{epsa0}
\epsilon_a^0=\frac{1}{3\pi} \frac{z_a\,z_b\,|u_a|^2\,M_a^2\,\sin\beta}{z_a^2\,t_a v^4 + 4\,|u_a|^4 M_a^2}\,,
\end{equation}
with $t_1=3$ and $t_2=2$. The matrix $\mathbf{P}^a$ is given by
\begin{align}
\mathbf{P}^a=\frac{(-1)^a}{2}\begin{pmatrix}
         -2(1+\varepsilon_1+\varepsilon_2) & \varepsilon_1-\varepsilon_2 & \varepsilon_2-\varepsilon_1\\
         \varepsilon_1-\varepsilon_2 &  4\left(\varepsilon_1+\varepsilon_2\, y_\mu^2/y_\tau^2\right) & 1+\varepsilon_1+\varepsilon_2\\
         \varepsilon_2-\varepsilon_1 & 1+\varepsilon_1+\varepsilon_2 & -4\left(\varepsilon_1+\varepsilon_2\, y_\mu^2/y_\tau^2\right)
       \end{pmatrix},
\end{align}
for case A, while
\begin{align}
\mathbf{P}^a=(-1)^a\left[\frac{1}{2} +\delta_{a2}\frac{v^4z_2^2(\varepsilon_1+\varepsilon_2)}{18M_2^2u_2^4 + 9v^4z_2^2}\right]\left(\begin{array}{rrr}
         -2 &\quad 0 &\quad 0\\
         0 &  0 & 1\\
         0 & 1 & 0
       \end{array}\right),
\end{align}
in case B. Obviously, in the TBM limit ($\varepsilon_{1,2}=0$), there is a unique matrix $\mathbf{P}^a$. In this case, the flavour structure of $\mathbf{P}^a$ dictates that the only allowed decay channels of $\Delta_a$ are into the $ee$ and $\mu\tau$ flavours. Once the VEV perturbations are introduced, new decay channels are opened in case A with the corresponding CP asymmetries suppressed by $\mathcal{O}(\varepsilon)$ factors.

\begin{figure}[t]
\begin{center}
\includegraphics[width=9cm]{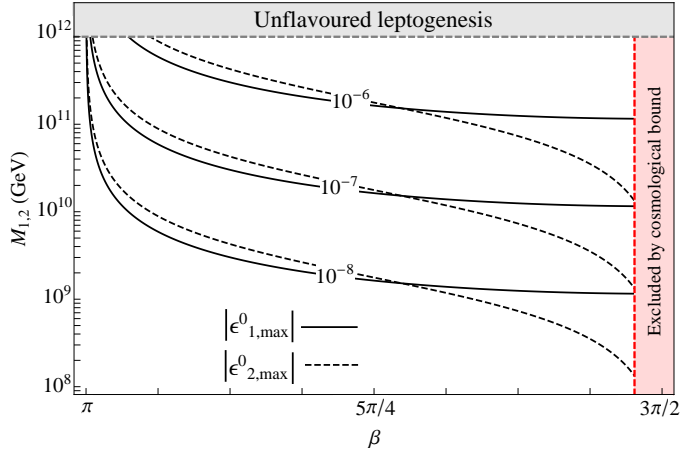}
\caption{\label{fig5} Contours of the maximum values $|\epsilon_{1,\text{max}}^0|$ (solid) and $|\epsilon_{2,\text{max}}^0|$ (dash), as given by Eq.~(\ref{epsmax}), in the $(\beta,M)$-plane.}
\end{center}
\end{figure}

Maximizing $\epsilon_a^0$ with respect to the VEV of the decaying scalar triplet $u_a$, one obtains
\begin{align}\label{epsmax}
    \epsilon^0_\text{1,max} \simeq \frac{M_1 \sqrt{\Delta m^2_{31}}}{12\sqrt{6} \pi v^2} \sin\beta, \quad \epsilon^0_\text{2,max} \simeq \frac{M_2 \sqrt{\Delta m^2_{31}}}{48 \pi v^2} \tan\beta\,.
\end{align}
In Fig.~\ref{fig5} we present the contours of $|\epsilon_{a,\text{max}}^0|$ in the $(\beta,M)$-plane. One can see that sufficiently large values of the CP asymmetries can be obtained in the flavoured regime. Clearly, a more rigorous study which accounts for washout effects would be necessary in order to estimate the final value of the baryon asymmetry. Such analysis is beyond the scope of the present work and will be presented elsewhere.

\section{Conclusions}

In this work we have presented an appealing scenario where spontaneous CP violation, leptonic mixing and  thermal leptogenesis are related in a simple way. We added a minimal particle content to the SM, namely, two Higgs triplets $\Delta_{1,2}$ and a complex scalar singlet $S$. In this framework, neutrinos acquire masses via the type II seesaw mechanism, implemented by the two scalar triplets. Furthermore, assuming that the CP symmetry is spontaneously broken at high energies by the complex VEV of the singlet scalar, the CP violation necessary for leptogenesis as well as non-vanishing leptonic CP violation at low energies are obtained. The model also exhibits a spontaneously broken $A_4 \times Z_4$ flavour symmetry which leads to a tribimaximal pattern in the leptonic mixing at low energies. Deviations to the TBM case were introduced in order to account for a non-zero value of the $\theta_{13}$ mixing angle, recently reported by the T2K, MINOS and Daya Bay experiments. Such deviations were achieved by perturbing the TBM vacuum alignments of the flavon fields. Two cases have been considered depending on whether the perturbation comes from the charged lepton (Case A) or neutrino (Case B) sector. We concluded that the latest oscillation data favours case B, leading to close-to-maximal Dirac CP-violation with $|J_{\rm CP}|\simeq 0.03-0.04$, which is large enough to be measured in a near future by neutrino oscillation experiments. The model also leads to large enough flavoured CP-asymmetries in the decays of the triplets into two leptons. Consequently, leptogenesis becomes viable due to the out-of-equilibrium decays of the triplets at temperatures below $10^{12}$~GeV. This could account for the observed value of the baryon asymmetry of the Universe.

\begin{acknowledgments}
We thank Mariam T\'{o}rtola for a private communication. The work of H.S. was supported by {\em Funda\c{c}\~{a}o para a Ci\^{e}ncia e a Tecnologia} (FCT, Portugal) under the Grant No. SFRH/BD/36994/2007. This work was supported by the projects CERN/FP/116328/2010, CFTP-FCT UNIT 777 and PTDC/FIS/098188/2008, which are partially funded through POCTI (FEDER).
\end{acknowledgments}

\end{document}